# Density functional calculation of the heats of formation of rare-earth orthophosphates

James R. Rustad, Sullivan Park Research Center. Corning, Inc., Corning, NY 14831

## Abstract

Electronic structure calculations are carried out to estimate the heats of formation of rare-earth orthophosphates from their oxides. The calculated heats of formation are systematically about 40 kJ/mol less exothermic than the measured values. Based on estimated corrections for zero-point energies and H(298.15)-H(0), the discrepancy is almost entirely electronic in origin. The decreasingly exothermic $\Delta H^f_{ox}$ with decreasing ionic radius (i.e. $LaPO_4$ more exothermic than $ScPO_4$) results from the higher charge localization on the oxide anion ($O^{2-}$) relative to the phosphate anion ($PO_4^{3-}$). The lattice energy, of course, becomes more negative with decreasing ionic radius for both oxide and phosphate phases, but does so more rapidly for the oxide, making the reaction less exothermic as ionic radius becomes smaller. This effect should carry over to $\Delta H^f_{ox}$ other oxyacids, such as silicates and sulfates.

## Introduction

The rare-earth (RE) phosphates are versatile refractory materials with a wide range of geological and technological applications (Ushakov et al., 2002). This paper reports the results of a series of density-functional electronic structure calculations of the thermodynamic properties of rare-earth phosphates. The purpose of the paper is to determine how well the calculated heats of formation from the oxides match measured values, and to compare the potential sources of error coming from lattice thermal contributions to the enthalpy, zero-point energies, and electronic energies.

This information is important for several reasons. First, it is only through studies of a systematic series of minerals that better density functionals for thermal properties can be developed. Studies focused only on individual phases cannot uncover systematic trends that are essential for making these improvements. Second, studies of interfacial properties of these minerals, in contact with both aqueous and silicate solvents, will require construction of potential functions (Pedone et al., 2006). While it is common to use structural and elastic properties to construct these potential functions, energetics of reactions such heats of formation from the oxides $\Delta H^f_{ox}$ (298.15K), (e.g. ½$Ln_2O_3$ + ½$P_2O_5$ = $LnPO_4$) are often not considered, mainly because the reported $\Delta H^f_{ox}$ will contain zero-point and enthalpic temperature corrections which cannot be made until vibrational analysis is carried out. Tables of electronic heats of formation at zero temperature, if available (and accurate, of course) would aid in the construction of transferable potential functions, which may eventually be capable of predicting thermodynamic properties of complex mixtures, such as glasses and melts. While interfacial phenomena in mineral-aqueous systems are well explored using simulation techniques, little work has been done on simulation of melt-crystal interfaces (Gurmani et al., 2010). Molecular processes at magma-mineral interfaces will require correct thermodynamics for reactions such as $2LaPO_4 = La_2O_3(s) + P_2O_5(melt)$. While obtaining the free energy of dissolution of $P_2O_5$ in a silicate melt will be out of reach of first-principles simulations methods for some time, it may be possible to obtain correct trends with parameterized potential functions, if they have been developed with attention paid to obtaining correct energetics. It is hoped that input on formation energies from first-principles methods will facilitate these efforts.

## Methods

Electronic structure calculations are carried out with density functional theory (DFT) using the Projector Augmented Wave (PAW) method (Bloechl, 1994) implemented in VASP 4.6.11 (Kresse and Furthmuller, 1996; Kresse and Hafner, 1993), with PAW pseudopotentials (Kresse and Joubert, 1999) constructed for the PBE exchange correlation functional (Perdew et al., 1996). The calculations on the lanthanide elements use the Ln_3 trivalent lanthanide pseudopotentials. These provide an implicit treatment of the *f* electron shell across the series of rare earth elements. Calculations for yttrium and scandium use the Y_sv and Sc_sv pseudopotentials. $LuPO_4$ failed to achieve SCF convergence, so this was eliminated from the study, along with Eu, Pm, and Yb. DFT+U calculations are carried for $Ce_2O_3$ and $CePO_4$ using the standard Ce PBE pseudopotential and a U value of 3 eV, which was found to be an optimal value both for $Ce_2O_3$ (Fabris et al., 2005; Loschen et al, 2007) and $CePO_4$ (Adelstein et al., 2011). The DFT+U calculations use the method of Dudarev et al (1998).

The cutoff energy was set to 500 eV for all systems. A gamma-centered reciprocal space grid was used for each system. The following k-point grid sizes were used: For the C-oxide, monazite, and xenotime structures, 7 x 7 x 7; for the A-oxide structures, 15x15x15; for *h*-$P_4O_{10}$ and *o'*-$P_2O_5$, 7x7x7. Structure optimizations were done with both volume and lattice vectors varying simultaneously. Final structures were run with no optimization to ensure that the absolute value of the pressure was below 0.1 GPa. Final configurations outside this range were optimized again until the pressure criterion was satisfied.

## Results and Discussion

### *Structure and Energetics of $RE_2O_3$ Oxides*

Tables 1, 2, and 3 give the optimized structural parameters and molar volume for some $RE_2O_3$ oxides, along with values obtained from x-ray measurements (Koehler and Wollan, 1953; Fert, 1962; Hase, 1963; Knop and Hartley, 1963; Boulesteix et al. 1971; Bourcherle et al., 1975; Saiki et al., 1985; Schiller, 1985; Bartos et al., 1993 Greis, 1994 ; Baldinozzi et al., 1998; Kuemmerle and Heger, 1999; Zhang et al., 2008). The large rare-earth oxides $La_2O_3$, $Ce_2O_3$, $Nd_2O_3$, and $Pr_2O_3$ are known to be most stable in the hexagonal A-type structure. Previous work (Wu et al. 2007) compared the electronic energies of a series of rare-earth oxides in the A and monoclinic B-type structure using the same method used here (PAW-PBE), but did not report results for the cubic C-type structure. An earlier PAW-GGA study (Hirosaki et al., 2003) gave structures but not energies for the C-type oxides. Table 4 gives the calculated energies for the suite of RE oxide compounds in the C-type structure as well as the A-type structure for La, Ce, Nd, and Pr, and the B-type structure for Sm and Gd. The results are very close to those previously calculated. The results show that PAW-PBE predicts (incorrectly) that La, Ce, Nd, Pr, and Sm are most stable in the C-type structure. This was already noted for $Ce_2O_3$ (Da Silva, 2007). As shown in Tables 1-3, calculated molar volumes are generally 2-3 percent greater than measured values, except for A-$Ce_2O_3$, which is 4.8 percent above the measured value. The PAW PBE+U calculations give improved calculations of the volume and lattice parameters for the A- and C-$Ce_2O_3$ phases. They are also reported to stabilize A-type $Ce_2O_3$ over C-type $Ce_2O_3$ (Da Silva, 2007) (DFT+U calculations were not done on C-$Ce_2O_3$ here).

*Structures and Energetics of $REPO_4$ Phosphates*

Tables 5 (xenotime) and 6 (monazite) give the optimized structural parameters and calculated energies per formula unit for the RE orthophosphates. Comparisons are made with the x-ray measurements (Ni et al., 1995; Milligan et al., 1982). The data overestimate volumes by close to 2% for the monazite-type $REPO_4$, and 2-3% for the xenotime $REPO_4$, except for $CePO_4$(m) which is overestimated by 4.6% with PBE and Ce_3 pseudopotential. The structure is improved slightly in the PBE+U treatment. In other respects, the structural calculations are unremarkable and in line with those of the oxides.

Calculated electronic energies for the $REPO_4$ are given in Table 7. La, Ce, Pr, and Nd are all most stable in the monazite structure, in agreement with observations (although actual measured values for $\Delta G$ of the monazite-xenotime polymorphs are not available for any of the $REPO_4$). For $NdPO_4$, the monazite and xenotime structures are nearly isoenergetic. For Tb-Tm, as well as Sc and Y, the xenotime structure is most stable, also in line with observations. $SmPO_4$ and $GdPO_4$ are also calculated to be more stable in the xenotime structure. The PBE+U calculations on $CePO_4$ give a low-energy antiferromagnetic structure almost isoenergetic with the ferromagnetic solution. The PBE+U treatment increases the stabilization of FM-$CePO_4$(m) relative to FM-$CePO_4$(x) by about 20 kJ/mol.

*Structures and Energetics of $P_2O_5$*

For calculation of the electronic energies of formation from the oxides, the total energy of $P_2O_5$ is needed. In the study of Ushakov et al. (2002), the $P_2O_5$ phase was the hexagonal *h*-$P_2O_5$. This is a molecular solid having discreet adamantane-like $P_4O_{10}$ molecules held together by van der Waals interactions which are not accounted for in the DFT. The omission of this contribution to the cohesive energy of $P_2O_5$ will produce a systematic error in the calculations, with the calculated $\Delta E_{e\ ox}^{f}$ being consistently too negative. It is known that the $h$-$P_2O_5$ phase is metastable with respect to a polymeric (i.e. non-molecular) orthorhombic $o'$-$P_2O_5$ compound (Greenwood and Earnshaw, 1985) in which molecular Van der Waals forces presumably play no role. As an upper bound to the required correction, it could be assumed that the total energies of $h$-$P_2O_5$ and $o'$-$P_2O_5$ are equal and the total energy of $o'$-$P_2O_5$ could be used as a surrogate for $h$-$P_2O_5$. The structural parameters and energies of both phases are given in Table 8 and compared with x-ray measurements (Cruickshank, 1964; Stachel et al. 1995). The calculated electronic energy of $o'$-$P_2O_5$ is 14.84 kJ/mol lower than $h$-$P_2O_5$.

*Formation of $REPO_4$ from $RE_2O_3$ and $P_2O_5$.*

In Table 9 electronic energies at zero temperature [$\Delta E_e{}^f{}_{ox}(0)=E_e(REPO_4)-1/2(E_e(P_2O_5)+E_e(RE_2O_3))$] for the reactions:

$$\tfrac{1}{2}\, P_2O_5 + \tfrac{1}{2}\, RE_2O_3 = REPO_4$$

are tabulated and compared with the measured heats of formation from the oxides at 298.15 K from Ushakov et al. (2002). Figure 1 shows the correlation between $\Delta H^f{}_{ox}(298.15)$ and the ionic radius noted by Ushakov et al. (2002), with the cube root of calculated molar volume of the cubic oxide serving as a convenient surrogate for the "computed" value of the ionic radius (this is simpler than, for example, trying to average the calculated RE-O distances). $GdPO_4$, in the monazite structure deviates most strongly from the correlation. $GdPO_4$ shows no anomalous predictions in the structural properties of either the $RE_2O_3$ or $REPO_4$ phases, and does not deviate strongly in the experimental correlation (Figure 3 in Ushakov et al. (2002)) so the reasons for the large deviation from the energy versus size correlation are not clear, except to note that using the more stable xenotime polymorph for $GdPO_4$(x) lies much closer to the correlation than $GdPO_4$(m). There is an indication of a separate, steeper trend for the monazite structures than the xenotime structures. Such a difference was not apparent on the experimental correlation. $\Delta H^f{}_{ox}(298.15)$ for $TbPO_4$(x) and $TbPO_4$(m) were measured within 3 kJ/mol of each other. This can be compared against the 23 kJ/mol difference in $\Delta E_e{}^f{}_{ox}(0)$ for these phases; so it can be concluded that the DFT calculations overestimate the energetic difference between the two polymorphs.

Figure 2 shows $\Delta H^f_{ox}(298.15)$-$\Delta E_e{}^f_{ox}(0)$ plotted again against the cube root of the volume of the cubic $RE_2O_3$ phase. The measured $\Delta H^f_{ox}(298.15)$ values lie fairly consistently about 40 kJ/mol lower than the calculated $\Delta E_e{}^f_{ox}(0)$, and exhibit no convincing trend with ionic radius, although again, the monazite polymorphs may comprise a trend of increasing $\Delta H^f_{ox}(298.15)$-$\Delta E_e{}^f_{ox}(0)$ with atomic number;. The average deviation would be increased to ~ 50 kJ/mol if *o'*-$P_2O_5$ were used instead of *h*-$P_2O_5$ as the $P_2O_5$ reference compound. The most negative deviations from the average (i.e. the $REPO_4$ phase is calculated to be thermodynamically less stable than average) occurs for $GdPO_4$ and $TbPO_4$. The most positive deviation is for $CePO_4$. Unlike $GdPO_4$, calculated structural properties of $CePO_4$ are anomalous. The predicted volume of $CePO_4$ is nearly 4.6 percent above x-ray measurements, in contrast to the 2-3 percent for the other $REPO_4$. The overestimation of the volume is 4.8 percent for the $Ce_2O_3$. Treatment of $CePO_4$ and $Ce_2O_3$ with PBE+U makes only a small correction (~ 3 kJ/mol) to the calculated $\Delta E_e{}^f_{ox}(0)$, so the effective RE_3 pseudopotentials are probably not the source of the error shown in Figure 2. The data in Figure 2 may suggest that the monazite $REPO_4$(m) phases lie on a different trend than the xenotime $REPO_4$(x) phases. If $LaPO_4$(m) is ignored, the difference between calculated and measured heats of formation for the monazite phases fall on a systematic trend with a negative slope (increasing atomic number implies increasingly negative deviation from experiment; in other words, as atomic number increases, the calculated heats of formation become less and less exothermic relative to the measured heats of formation).

*Thermal Corrections to* $\Delta E_e{}^f_{ox}(0)$

Direct comparison of the calculated $\Delta E_e{}^f_{ox}(0)$) for the formation of the $REPO_4$ from oxides with experimental $\Delta H$(298 K) requires, at a minimum, knowledge of the differential zero-point energies $\Delta ZPE=ZPE(LnPO_4)-1/2[ZPE(P_2O_5)+ZPE(Ln_2O_3)]$ and the differential enthalpies $\Delta\int C_p dT= \int(C_p(LnPO_4)-1/2[C_p(Ln_2O_3)+C_p(P_2O_5)]dT$ between ($P_2O_5$ + $Ln_2O_3$) and $LnPO_4$. These could, in principle, be computed from DFT for all phases, but would require a heavy investment of computational resources. Here, an empirical approach is taken to estimate the likely magnitude of the thermal contributions. There will also be Schottky-type contributions for some of the phases (Westrum, 1985), but these cannot be responsible for the overall 40 kJ/mol differential between the measured and calculated enthalpies, as they should be close to zero for compounds with no crystal field stabilization (Y, Sc, La, Gd).

The $\Delta \int Cp\, dT$ can be summed from low-temperature heat capacities of the $RE_2O_3$, $P_2O_5$, and $REPO_4$ phases. Low-temperature heat capacities for the $RE_2O_3$ and $REPO_4$ phases (Goldstein et al., 1959; Justice and Westrum, 1963; Weller and King, 1963; Justice et al, 1969; Gavrichev et al., 1993; Gruber et al., 2002) are given in Table 10. Low-temperature heat capacities have been measured for $YPO_4$ (Gavrichev et al., 2010a), $ScPO_4$(Gavrichev et al., 2010b), $LaPO_4$ (Gavrichev et al. 2008), and $LuPO_4$ (Gavrichev et al, 2006). The low-temperature heat capacity of $P_2O_5$ has been measured at 16.98 kJ/mol (Andon et al., 1963); first principles electronic structure calculations yield estimates of H(298.15)-H(0)=16.37 kJ/mol (Rustad, 2011). Taking typical values of ~20 kJ/mol for $RE_2O_3$, ~17 kJ/mol for $REPO_4$, and 16 kJ/mol for $P_2O_5$, the thermal enthalpy correction to $\Delta E_e{}^f{}_{ox}(0)$ is close to zero (-1 kJ/mol), and cannot account for the systematic difference between $\Delta H$(298 K) and $\Delta E_e{}^f{}_{ox}(0)$.

Estimates for ZPEs can be made empirically from knowledge of the infrared and Raman vibrational spectra for each of the phases. Uncertainties related to the deconvolution of multicomponent peaks, knowledge of vibrational degeneracies, and anharmonic effects contribute to inaccuracies in these estimates. A normal coordinate analysis has been carried for A-type $(La\text{-}Pr\text{-}Nd)_2O_3$ (Gonipath and Brown, 1982). Infrared and Raman spectra have been measured for $LnPO_4$ (Ln=La, Ce, Pr, Nd, Sm, Eu, Gd) (Silva et al, 2006). Infrared and Raman spectra have been obtained for $h$-$P_4O_{10}$ (Gilliam et al., 2003), however the ZPE for $P_2O_5$ (70.6 kJ/mol) is taken from a recent theoretical calculation of the vibrational spectrum of this phase (Rustad, 2011) because of some revisions in the interpretation of the measured spectrum indicated by the calculations. The vibrational frequencies and ZPE estimates are given in Tables 11 and 12. Again, taking typical values, an estimate for $\Delta$ZPE is 58 kJ/mol – ½(21.5 -70.6). This yields a +12 kJ/mol correction to the calculated $\Delta E_e{}^f{}_{ox}(0)$, in the opposite direction of the correction needed to bring the calculated $\Delta E_e{}^f{}_{ox}(0)$ in consonance with measured values of ΔH(298 K); in other words, the calculated heats become ~12 kJ/mol *less* exothermic by accounting for the ZPE. The calculated heats are already not exothermic enough.

To check these estimates, first principles calculations of the vibrational spectrum of $YPO_4$ and cubic $Y_2O_3$ were carried out using the CASTEP planewave-pseudopotential code (Clark et al., 2005), as implemented within Materials Studio of Accelrys, Inc. A plane-wave cutoff of 750 eV and norm-conserving pseudopotentials (Lin et al., 1993; Lee, 1996) were used to do the calculations. The phonons were calculated with the linear-response method (Refson et al., 2006). For $Y_2O_3$ and $YPO_4$ the norm-conserving pseudopotentials give structures that compare well with the VASP calculations and experiment ($Y_2O_3$: a=10.594 Å; $YPO_4$ a=6.957 Å, c=5.964 Å) Table 13 gives the thermodynamic properties of $Y_2O_3$, $YPO_4$, and $P_2O_5$. The calculated $\Delta E_{e}{}^{f}{}_{ox}(0)$ is, remarkably, within 1 kJ/mol of the value calculated with VASP. The calculated corrections to obtain $\Delta H^{f}_{ox}(298.15)$ from $\Delta E_{e}{}^{f}{}_{ox}(0)$ from both the zero-point energy and the heat capacity are small, on the order of 1 kJ/mol or less. The main discrepancy with the empirical estimate is the calculated ZPE of the oxide. The calculated value for $Y_2O_3$ (33.5 kJ/mol) is higher than the value estimated from a force field parameterized against the measured vibrational spectra of A-type $(La, Pr, Nd)_2O_3$ given in Gopinath and Brown (1982). These are, of course, different structures. Calculated frequencies at q=(0,0,0) are given for $YPO_4$ and $Y_2O_3$ in Tables 14 and 15, and the phonon density of states for $YPO_4$(x) and c-$Y_2O_3$ are given in Figures 3 and 4. For $YPO_4$(x), there is good agreement between the vibrational frequencies and the Raman spectrum for xenotime reported in (http://www.ens-lyon.fr/LST/Raman). For $Y_2O_3$ there appear to be contributions at significantly higher frequencies than indicated in the measured spectrum (Repelin et al., 1995). It seems reasonable to conclude that the ΔZPE correction probably lies somewhere between 0 and +12 kJ/mol.

*Reactivity Trends*

The overall trend of increasing $\Delta H^{f}_{ox}$ with ionic radius observed by Ushakov et al., (2002) Reference, and reproduced here using electronic structure calculations, is easy to understand. The electrostatic energies of the oxides and phosphate phases were computed in the ionic model assuming formal charges of +3 for the REE, -2 for oxygen, and +5 for phosphorous. Figure 5 shows the relationship between the electrostatic energy and the ionic radius for the oxide and phosphate phases. To make the comparison easier, the lattice energy of *h*-$P_2O_5$ has been added to the lattice energy of the oxide phases and the sum is multiplied by ½. The difference indicated in the figure is therefore the electrostatic energy of formation (this quantity is highly exothermic because it neglects cation-anion repulsion). The reason for the observed trend is simply that, due to the localized charge on the oxide anions relative to the phosphate anions, the lattice energies of the oxide phases are a stronger function of ionic radius than the phosphates (compare the slopes of 0.82/2 eV/pm for the oxides and 0.34 eV/pm for the phosphates). In other words, the lattice energy becomes more negative for both phases as ionic radius decreases, but does so more rapidly for oxide phase than the phosphate phase because the anionic charge is more localized in the oxide. This type of trend governs the heats of formation of other oxyacid compounds as well, such as sulfates, carbonates, and silicates, as shown in Figure 6, with $\Delta H^f_{ox}$ taken from the FACTSAGE thermodynamic database.

**Conclusions**

Density functional electronic structure calculations have been carried out on Sc, Y and RE orthophosphates and oxides, and *h*- and *o'*-$P_2O_5$ polymorphs to calculate the electronic heats of formation $\Delta E_e{}^f_{ox}(0)$. Calculated heats of formation are systematically approximately 40 kJ/mol less exothermic than measured $\Delta H^f_{ox}(298.15)$. The systematic error is not uniform with large deviations for Gd(m) (-64.7 kJ/mol) and Ce(m) (-27.6 kJ/mol). For systems where low-temperature heat capacity data are available, H(298.15)-H(0) corrections are estimated to be less than 1 kJ/mol. Empirical estimates for ZPE of orthophosphate and oxide phases are nearly independent of atomic number, and give a 0-+12 kJ/mol correction to the $\Delta E_e{}^f_{ox}(0)$. Ab initio calculation of the zero-point correction for formation of $YPO_4$, based on the computed vibrational spectrum of $YPO_4$ and $Y_2O_3$, gives ΔZPE and Δ[H(298.15)-H(0)] close to zero. Thus the origin of the 40 kJ/mol discrepancy between the measured values of $\Delta H^f_{ox}(298.15)$ and calculated values of $\Delta E_e{}^f_{ox}(0)$ appears to be electronic in origin. DFT+U calculations for $Ce_2O_3$ and $CePO_4$ make only a 3 kJ/mol correction, indicating that the systematic difference cannot be corrected by including electron correlation at the DFT+U level. The observed correlation between atomic number and heat of formation results from the lanthanide contraction and the localized charge on the oxide anion relative to the phosphate anion and can be reproduced from the simplest ionic model.

**Figure Captions**

Figure 1. Correlation between $\Delta E_e{}^f{}_{ox}(0)$ and the cube root of the computed volume of the $RE_2O_3$ phase. (m) monazite, (x) xenotime.

Figure 2. Correlation between the difference of the measured heat of formation and the computed electronic energy [$\Delta H^f_{ox}(298.15)$- $\Delta E_e{}^f{}_{ox}(0)$] and the cube root of the computed volume of the cubic $RE_2O$.

Figure 3. Phonon density of states computed for $YPO_4$-xenotime.

Figure 4. Phonon density of states computed for cubic $Y_2O_3$.

Figure 5. Electrostatic lattice energies E as a function of ionic radius for RE oxide and phosphate phases. The lattice energy of *h*-$P_2O_5$ has been added to the oxides and the sum multiplied by ½.

Figure 6. Correlation between heat of formation from the oxides and ionic radius for carbonate ($MCO_3$), orthosilicate ($M_2SiO_4$), and sulfate($MSO_4$) compounds with divalent anions. Data taken from FACTSAGE thermodynamic database.

Table 1. Structural Parameters for hexagonal A-type $RE_2O_3$ (volumes in $Å^3$/formula unit)

| | Calculated | | | | | | | | | Measured | | |
|---|---|---|---|---|---|---|---|---|---|---|---|---|
| | This work | | | Hirosaki et al., (2003) | | | Wu et al. (2007) | | | | | |
| | a (Å) | c (Å) | V($Å^3$) | a (Å) | c (Å) | V($Å^3$) | a (Å) | c (Å) | V($Å^3$) | a (Å) | c (Å) | V($Å^3$) |
| $La_2O_3$[a] | 3.936 | 6.200 | 83.20 | 3.936 | 6.166 | 82.73 | 3.938 | 6.173 | 82.90 | 3.940 | 6.13 | 82.410 |
| $Ce_2O_3$[b] | 3.944 | 6.180 | 83.26 | 3.941 | 6.182 | 83.14 | 3.944 | 6.191 | 83.38 | 3.891 | 6.059 | 79.440 |
| $Ce_2O_3$ (PBE+U) AFM | 3.897 | 6.229 | 81.91 | | | | | | | | | |
| $Ce_2O_3$ (PBE+U) FM | 3.892 | 6.224 | 81.66 | | | | | | | | | |
| $Pr_2O_3$[c] | 3.896 | 6.135 | 80.64 | 3.895 | 6.126 | 80.50 | 3.899 | 6.135 | 80.76 | 3.859 | 6.0131 | 77.550 |
| $Nd_2O_3$[d] | 3.867 | 6.082 | 78.77 | 3.859 | 6.072 | 78.30 | 3.859 | 6.090 | 78.54 | 3.831 | 5.999 | 76.250 |

[a]Koehler and Wollan, 1953;[b]Schiller, 1985; [c]Hase, 1963; [d]Boucherle and Schweizer, 1975

Table 2. Structural Parameters for monoclinic B-type $RE_2O_3$ (volume in $Å^3$/formula unit)

| | a (Å) | b (Å) | c (Å) | β | V($Å^3$) |
|---|---|---|---|---|---|
| calc (this work) | | | | | |
| $Sm_2O_3$ | 14.384 | 3.631 | 8.916 | 100.26 | 76.36 |
| $Gd_2O_3$ | 14.177 | 3.565 | 8.770 | 100.30 | 72.68 |
| calc[a] | | | | | |
| $Sm_2O_3$ | 14.381 | 3.635 | 8.911 | 100.15 | 76.41 |
| $Gd_2O_3$ | 14.195 | 3.566 | 8.770 | 100.18 | 72.80 |
| measured | | | | | |
| $Sm_2O_3$[b] | 14.198 | 3.627 | 8.856 | 99.99 | 74.66 |
| $Gd_2O_3$[c] | 14.032 | 3.583 | 8.742 | 100.13 | 72.11 |

[a]Wu et al., 2007; [b]Boulestix et al, 1971; [c]Zhang et al., 2008

Table 3. Structural Parameters for cubic C-type $RE_2O_3$ (volume in $Å^3$/formula unit)

| | calculated | | | | measured | |
|---|---|---|---|---|---|---|
| | This work | | Hirosaki et al., (2003) | | | |
| | a (Å) | V($Å^3$) | a (Å) | V($Å^3$) | a (Å) | V($Å^3$) |
| $Sc_2O_3$ | 9.911 | 60.85 | | | 9.846[a] | 59.66 |
| $Y_2O_3$ | 10.701 | 76.58 | | | 10.596[b] | 74.36 |
| $La_2O_3$ | 11.387 | 92.29 | 11.392 | 92.40 | | |
| $Ce_2O_3$ | 11.414 | 92.94 | 11.410 | 92.84 | 11.111[c] | 85.73 |
| $Pr_2O_3$ | 11.290 | 89.94 | 11.288 | 89.89 | | |
| $Nd_2O_3$ | 11.178 | 87.30 | 11.176 | 87.24 | | |
| $Sm_2O_3$ | 10.998 | 83.14 | 10.995 | 83.07 | 10.930[d] | 81.61 |
| $Gd_2O_3$ | 10.819 | 79.16 | 10.812 | 78.99 | 10.790[d] | 78.51 |
| $Tb_2O_3$ | 10.744 | 77.50 | | | 10.729[e] | 77.10 |
| $Dy_2O_3$ | 10.675 | 76.02 | 10.670 | 75.92 | 10.670[d] | 75.92 |
| $Ho_2O_3$ | 10.609 | 74.63 | 10.605 | 74.54 | 10.580[d] | 74.02 |
| $Er_2O_3$ | 10.544 | 73.26 | 10.544 | 73.26 | 10.548[f] | 73.35 |
| $Tm_2O_3$ | 10.472 | 71.77 | | | 10.480[g] | 71.94 |

[a]Knop and Hartley, 1963; [b]Baldinozzi et al., 1998; [c]Kuemmerle and Heger, 1999; [d]Bartos et al., 1993; [e]Saiki et al., 1985; [f]Fert, 1962; [g]Hase, 1963

Table 4. Energies of $RE_2O_3$ phases (eV/formula unit).

| | C-type | A-type | B-type | A-type[a] | B-type[a] |
|---|---|---|---|---|---|
| $Sc_2O_3$ | -45.2887 | | | -44.6351 | -44.9478 |
| $Y_2O_3$ | -45.5652 | | | -45.2299 | -45.3291 |
| $La_2O_3$ | -42.0219 | -41.9007 | | -41.9079 | -41.8576 |
| $Ce_2O_3$ | -40.7455 | -40.6337 | | -40.6376 | -40.6027 |
| $Ce_2O_3$ (PBE+U) AFM | | -41.6814 | | | |
| $Ce_2O_3$ (PBE+U) FM | | -41.6791 | | | |
| $Pr_2O_3$ | -41.0188 | -40.8856 | | -40.8945 | -40.8744 |
| $Nd_2O_3$ | -41.2258 | -41.0714 | | -41.0803 | -41.0746 |
| $Sm_2O_3$ | -41.5301 | | -41.3677 | -41.3363 | -41.362 |
| $Gd_2O_3$ | -41.8636 | | -41.6746 | -41.6172 | -41.6718 |
| $Tb_2O_3$ | -41.9675 | | | -41.7017 | -41.7655 |
| $Dy_2O_3$ | -42.0428 | | | -41.7398 | -41.8283 |
| $Ho_2O_3$ | -42.1142 | | | -41.8004 | -41.8883 |
| $Er_2O_3$ | -42.2007 | | | -41.8377 | -41.9627 |
| $Tm_2O_3$ | -42.1943 | | | -41.7963 | -41.9437 |

[a]Wu et al., 2007

Table 5. Structural parameters for $REPO_4$ in the xenotime (zircon) structure (volume in $Å^3$/formula unit)

| | calculated | | | measured | | |
|---|---|---|---|---|---|---|
| | a (Å) | c(Å) | V($Å^3$) | a (Å) | c(Å) | V($Å^3$) |
| $ScPO_4$[a] | 6.661 | 5.837 | 64.740 | 6.574 | 5.791 | 62.568 |
| $YPO_4$[a] | 6.976 | 6.079 | 73.950 | 6.895 | 6.028 | 71.633 |
| $TbPO_4$[b] | 6.984 | 6.086 | 74.205 | 6.931 | 6.061 | 72.784 |
| $DyPO_4$[b] | 6.953 | 6.066 | 73.328 | 6.905 | 6.038 | 71.981 |
| $HoPO_4$[b] | 6.934 | 6.050 | 72.710 | 6.877 | 6.018 | 71.154 |
| $ErPO_4$[b] | 6.905 | 6.026 | 71.823 | 6.851 | 5.997 | 70.363 |
| $TmPO_4$[b] | 6.880 | 5.998 | 70.968 | 6.829 | 5.980 | 69.726 |
| $LaPO_4$ | 7.274 | 6.347 | 83.945 | | | |
| $CePO_4$ | 7.261 | 6.342 | 83.575 | | | |
| $CePO_4$ (PBE+U[c]) | 7.233 | 6.314 | 82.62 | | | |
| $PrPO_4$ | 7.209 | 6.293 | 81.768 | | | |
| $NdPO_4$ | 7.178 | 6.247 | 80.473 | | | |
| $SmPO_4$ | 7.089 | 6.187 | 77.728 | | | |
| $GdPO_4$ | 7.015 | 6.116 | 75.248 | | | |

[a]Milligan et al, 1982; [b]Ni et al., 1995; [c]ferromagnetic electronic state

Table 6. Structural parameters for $REPO_4$ in the monazite structure. (volume in $Å^3$/formula unit)

| | calculated | | | | | measured[a] | | | | |
|---|---|---|---|---|---|---|---|---|---|---|
| | a (Å) | b(Å) | c(Å) | β (deg) | V($Å^3$) | a (Å) | b(Å) | c(Å) | β (deg) | V($Å^3$) |
| $LaPO_4$ | 6.932 | 7.142 | 6.543 | 103.6 | 78.71 | 6.831 | 7.071 | 6.503 | 103.3 | 76.43 |
| $CePO_4$ | 6.916 | 7.135 | 6.535 | 103.5 | 78.40 | 6.788 | 7.016 | 6.465 | 103.4 | 74.98 |
| AFM1[b] | 6.900 | 7.110 | 6.510 | 103.5 | 77.62 | | | | | |
| AFM2 | 6.883 | 7.099 | 6.522 | 103.6 | 77.44 | | | | | |
| AFM3 | 6.908 | 7.113 | 6.515 | 103.5 | 77.82 | | | | | |
| FM[c] | 6.899 | 7.105 | 6.518 | 103.6 | 77.64 | | | | | |
| $PrPO_4$ | 6.873 | 7.079 | 6.485 | 103.8 | 76.61 | 6.760 | 6.981 | 6.434 | 103.5 | 73.80 |
| $NdPO_4$ | 6.840 | 7.040 | 6.452 | 103.8 | 75.42 | 6.735 | 6.950 | 6.405 | 103.7 | 72.83 |
| $SmPO_4$ | 6.803 | 6.961 | 6.394 | 104.2 | 73.38 | 6.682 | 6.888 | 6.365 | 103.9 | 71.10 |
| $GdPO_4$ | 6.713 | 6.887 | 6.358 | 104.2 | 71.24 | 6.644 | 6.841 | 6.328 | 104.0 | 69.78 |
| $TbPO_4$ | 6.687 | 6.858 | 6.337 | 104.2 | 70.42 | | | | | |

[a]Ni et al., 1995; [b]antiferromagnetic electronic states;[c]ferromagnetic electronic state

Table 7. Calculated Energies of $REPO_4$ (eV/formula unit).

| | x-type | m-type |
|---|---|---|
| $ScPO_4$(x) | -48.845 | |
| $YPO_4$(x) | -49.760 | |
| $LaPO_4$(m) | -48.215 | -48.337 |
| $CePO_4$(m) | -47.679 | -47.805 |
| $CePO_4$ (PBE+U) AFM1 | | -48.280 |
| $CePO_4$ (PBE+U) AFM2 | | -48.304 |
| $CePO_4$ (PBE+U) AFM3 | | -48.251 |
| $CePO_4$ (PBE+U) FM | -47.970 | -48.302 |
| $PrPO_4$(m) | -47.779 | -47.843 |
| $NdPO_4$(m) | -47.846 | -47.854 |
| $SmPO_4$(m) | -47.923 | -47.835 |
| $GdPO_4$(m) | -48.013 | -47.819 |
| $TbPO_4$(x) | -48.027 | -47.788 |
| $DyPO_4$(x) | -48.025 | |
| $HoPO_4$(x) | -48.022 | |
| $ErPO_4$(x) | -48.031 | |
| $TmPO_4$(x) | -47.991 | |

Table 8. Structural parameters and energies of $P_2O_5$ phases.

| | (a,b,c)(Å) | (a,b,g) deg | | V(Å$^3$) | E (eV/formula unit) |
|---|---|---|---|---|---|
| h-$P_2O_5$ | 7.592 | 87.6 | | 109.1 | -48.9743 |
| measured[a] | 7.43 | 87 | | 102.1 | |
| | a(Å) | b(Å) | c(Å) | | |
| o'-$P_2O_5$ | 9.534 | 4.952 | 7.308 | 345.0 | -49.1281 |
| measured[b] | 9.139 | 4.89 | 7.162 | 320.1 | |

[a]Cruickshank, 1964

[b]Stachel et al., 1995

Table 9. Measured vs. calculated electronic energy (kJ/mol)

| | $\Delta H^{f}_{ox}(298.15)$ kJ/mol[a] | $\Delta E_{e}{}^{f}{}_{ox}(0)$ kJ/mol | $\Delta H^{f}_{ox}(298.15)$- $\Delta E_{e}{}^{f}{}_{ox}(0)$ kJ/mol |
|---|---|---|---|
| $YPO_4$ (x) | -282.6 | -240.2 | -42.4 |
| $LaPO_4$m) | -321.4 | -279.7 | -41.7 |
| $LaPO_4$ (x) | | -268.0 | -53.4 |
| $CePO_4$ (m) | -317.2 | -289.6 | -27.6 |
| PBE+U AFM1 | | -287.1 | -30.1 |
| PBE+U AFM2 | | -284.8 | -32.4 |
| PBE+U AFM3 | | -282.0 | -35.2 |
| PBE+U FM | | -287.0 | -30.2 |
| $CePO_4$ (x) | | -277.4 | -39.8 |
| $PrPO_4$ (m) | -312.2 | -281.1 | -31.1 |
| $PrPO_4$ (x) | | -274.8 | -37.4 |
| $NdPO_4$ (m) | -312.0 | -273.2 | -38.8 |
| $NdPO_4$x) | | -272.3 | -39.7 |
| $SmPO_4$ (m) | -301.8 | -257.1 | -44.7 |
| $SmPO_4$ (x) | | -265.5 | -36.3 |
| $GdPO_4$ (m) | -296.2 | -231.6 | -64.6 |
| $GdPO_4$ (x) | | -250.3 | -45.9 |
| $TbPO_4$ (x) | -286.1 | -246.6 | -39.5 |
| $TbPO_4$ (m) | -283.5 | -223.5 | -60.0 |
| $DyPO_4$ (x) | -283.9 | -242.8 | -41.1 |
| $HoPO_4$ (x) | -278.8 | -239.1 | -39.7 |
| $ErPO_4$ (x) | -275.6 | -235.7 | -39.9 |
| $TmPO_4$ (x) | -268.0 | -232.2 | -35.8 |
| $ScPO_4$ (x) | -209.8 | -165.4 | -44.4 |

[a]Ushakov et al., 2002

Table 10. Enthalpy correction to standard temperature H(298.15)-H(0) for $RE_2O_3$ phases (kJ/mol)

| | $RE_2O_3$ | $REPO_4$ |
|---|---|---|
| Sc | 13.845[a] | 14.934[g] |
| Y | 16.800[b] | 15.944[h] |
| La | 19.842[c] | 17.440[i] |
| Ce | 21.479[c] | |
| Pr | 22.734[c] | |
| Nd | 20.892[c] | |
| Sm | 21.008[d] | |
| Gd | 18.510[d] | |
| Dy | 21.025[e] | |
| Ho | 20.958[e] | |
| Er | 19.995[e] | |
| Tm | 20.887[f] | |
| Lu | 17.539[f] | 16.430[j] |

[a]Weller and King, 1963;[b]Gavrichev et al., 1993;[c]Gruber et al., 2002;[d]Justice and Westrum, 1963;[e]Westrum and Justice, 1963;[f]Justice et al., 1969; [g]Gavrichev et al, 2010a;[h]Gavrichev et al., 2010b;[i]Gavrichev et al. 2008;[j]Gavrichev et al., 2006

Table 11. Vibrational frequencies ($cm^{-1}$) with $h\nu/2$ contribution to zero point energy (kJ/mol) for monazite-type $LnPO_4$.[a]

| $LaPO_4$ | | $CePO_4$ | | $PrPO_4$ | | $NdPO_4$ | | $SmPO_4$ | | $EuPO_4$ | | $GdPO_4$ | |
|---|---|---|---|---|---|---|---|---|---|---|---|---|---|
| 90 | 0.5 | 88 | 0.5 | 90 | 0.5 | 89 | 0.5 | 88 | 0.5 | 87 | 0.5 | 87 | 0.5 |
| 100 | 0.6 | 100 | 0.6 | 105 | 0.6 | 106 | 0.6 | 107 | 0.6 | 108 | 0.6 | 108 | 0.6 |
| 151 | 0.9 | 152 | 0.9 | 153 | 0.9 | 154 | 0.9 | 155 | 0.9 | 156 | 0.9 | 158 | 0.9 |
| 157 | 0.9 | 158 | 0.9 | 160 | 1.0 | 160 | 1.0 | 159 | 1.0 | 160 | 1.0 | 162 | 1.0 |
| 170 | 1.0 | 172 | 1.0 | 176 | 1.1 | 175 | 1.0 | 177 | 1.1 | 175 | 1.0 | 178 | 1.1 |
| 184 | 1.1 | 183 | 1.1 | 182 | 1.1 | 183 | 1.1 | 185 | 1.1 | 189 | 1.1 | 192 | 1.1 |
| 219 | 1.3 | 219 | 1.3 | 225 | 1.3 | 228 | 1.4 | 231 | 1.4 | 234 | 1.4 | 236 | 1.4 |
| 226 | 1.4 | 227 | 1.4 | 233 | 1.4 | 236 | 1.4 | 243 | 1.5 | 243 | 1.5 | 247 | 1.5 |
| 258 | 1.5 | 254 | 1.5 | 260 | 1.6 | 264 | 1.6 | 265 | 1.6 | 265 | 1.6 | 268 | 1.6 |
| 275 | 3.3 | 270 | 1.6 | 282 | 1.7 | 291 | 1.7 | 293 | 1.8 | 298 | 1.8 | 302 | 1.8 |
| 396 | 2.4 | 396 | 2.4 | 299 | 1.8 | 398 | 2.4 | 404 | 2.4 | 404 | 2.4 | 406 | 2.4 |
| 413 | 2.5 | 414 | 2.5 | 417 | 5.0 | 419 | 2.5 | 424 | 2.5 | 425 | 2.5 | 428 | 2.6 |
| 466[b] | 5.6 | 467 | 5.6 | 470 | 5.6 | 471 | 5.6 | 474 | 5.7 | 472 | 5.6 | 478 | 5.7 |
| 620[c] | 11.1 | 618 | 11.1 | 623 | 11.2 | 625 | 11.2 | 629 | 11.3 | 631 | 11.3 | 634 | 11.4 |
| 968[d] | 5.8 | 970 | 5.8 | 975 | 5.8 | 977 | 5.8 | 983 | 5.9 | 990 | 5.9 | 988 | 5.9 |
| 1054[e] | 18.9 | 1054 | 18.9 | 1058 | 19.0 | 1061 | 19.0 | 1065 | 19.1 | 1069 | 19.2 | 1072 | 19.2 |
| | 58.8[f] | | 57.1[f] | | 59.6[f] | | 57.9[f] | | 58.3[f] | | 58.5[f] | | 58.8[f] |

[a]data taken from Silva et al., 2006.

[b]$PO_4$-$\nu_2$ (multiplicity 2), [c]$PO_4$-$\nu_4$ (multiplicity 3), [d]$PO_4$-$\nu_1$ (multiplicity 1), [e]$PO_4$-$\nu_3$ (multiplicity 3),

[f] total zero-point energy ($\Sigma h\nu_i/2$)

Table 12. Vibrational frequencies ($cm^{-1}$) with hν/2 contribution to zero point energy (kJ/mol) for A-type $Ln_2O_3$ oxide phases.[a]

| | $La_2O_3$ | | $Pr_2O_3$ | | $Nd_2O_3$ | |
|---|---|---|---|---|---|---|
| $E_g$ | 413 | 4.9 | 415 | 5.0 | 434 | 5.2 |
| $A_{1g}$ | 400 | 2.4 | 404 | 2.4 | 422 | 2.5 |
| $E_u$ | 408 | 4.9 | 409 | 4.9 | 415 | 5.0 |
| $A_{2a}$ | 404 | 2.4 | 406 | 2.4 | 407 | 2.4 |
| $E_u$ | 243 | 2.9 | 264 | 3.2 | 232 | 2.8 |
| $A_{2u}$ | 256 | 1.5 | 258 | 1.5 | 223 | 1.3 |
| $A_{1g}$ | 191 | 1.1 | 189 | 1.1 | 190 | 1.1 |
| $E_g$ | 99 | 1.2 | 99 | 1.2 | 100 | 1.2 |
| | | 21.4[b] | | 21.7[b] | | 21.6[b] |

[a]data taken from Gopinath and Brown, 1982.
[b]total zero-point energy ($\Sigma h\nu_i/2$)

Table 13. Thermodynamic properties obtained with CASTEP

| | $E_e$ | ZPE [kJ $mol^{-1}$] | Cv [J $mol^{-1}$ $K^{-1}$] | H(298.15)-H(0) [kJ $mol^{-1}$] | S(298.15) [J $K^{-1}$ $mol^{-1}$] |
|---|---|---|---|---|---|
| $YPO_4$ calc (this work) | -1957.54 eV | 53.14 | 96.84 J $mol^{-1}$ $K^{-1}$ (Cv) | 15.001 | 87.28 |
| exp[a] | | - | 99.27 J $mol^{-1}$ $K^{-1}$ (Cp) | 15.994 | 93.86 |
| $Y_2O_3$ calc (this work) | -1393.97 eV | 33.45 | 93.66 | 14.760 | 85.92 |
| exp[b] | | - | 103.4 | 16.800 | 98.96 |
| *h*-$P_2O_5$ calc[c] | -2516.15 eV | 70.65 | 103.1 J $K^{-1}$ $mol^{-1}$ | 16.370 | 106.45 |
| $\Delta E_{e\ ox}^{f}(0)$ | -239.28 kJ/mol | | | | |
| Correction to H calc | | 1.09 | | -0.56 | |
| *exp*[d] | -282.6 kJ/mol | | | -0.59 | |

[a]Gavrichev et al., 2010; [b]Gavrichev et al., 1993; [c]Rustad, 2011; [d]Ushakov et al., 2002.

Table 14. Computed vibrational modes at q=(0,0,0) for $Y_2O_3$

| $\nu(cm^{-1})$ | g[a] | IR Int[b] | IR[c] | RM[b] |
|---|---|---|---|---|
| 114.4 | 1 | 0.00 | N | N |
| 122.0 | 3 | 0.04 | Y | N |
| 132.1 | 3 | 0.00 | N | Y |
| 135.7 | 2 | 0.00 | N | N |
| 141.1 | 3 | 0.00 | N | Y |
| 159.0 | 3 | 0.01 | Y | N |
| 166.6 | 1 | 0.00 | N | Y |
| 177.2 | 3 | 0.23 | Y | N |
| 187.5 | 3 | 0.17 | Y | N |
| 193.0 | 3 | 0.00 | N | Y |
| 207.1 | 2 | 0.00 | N | Y |
| 207.9 | 3 | 0.06 | Y | N |
| 241.7 | 1 | 0.00 | N | N |
| 253.2 | 3 | 0.00 | N | Y |
| 257.3 | 3 | 0.00 | N | Y |
| 260.7 | 2 | 0.00 | N | N |
| 261.6 | 3 | 1.08 | Y | N |
| 320.2 | 3 | 0.00 | Y | N |
| 378.6 | 3 | 88.12 | Y | N |
| 385.1 | 3 | 0.00 | N | Y |
| 393.5 | 3 | 2.04 | Y | N |
| 394.9 | 1 | 0.00 | N | N |
| 397.1 | 2 | 0.00 | N | Y |
| 398.5 | 3 | 0.00 | N | Y |
| 412.5 | 3 | 51.24 | Y | N |
| 419.7 | 1 | 0.00 | N | Y |
| 420.6 | 2 | 0.00 | N | N |
| 425.5 | 3 | 1.70 | Y | N |
| 435.1 | 3 | 0.00 | N | Y |
| 448.6 | 3 | 100.11 | Y | N |
| 452.3 | 3 | 0.00 | N | Y |
| 463.4 | 2 | 0.00 | N | Y |
| 470.6 | 2 | 0.00 | N | N |
| 471.6 | 3 | 0.00 | N | Y |
| 476.5 | 3 | 3.71 | Y | N |
| 506.0 | 1 | 0.00 | N | Y |
| 514.7 | 3 | 0.00 | N | Y |
| 518.5 | 1 | 0.00 | N | N |
| 551.8 | 3 | 4.12 | Y | N |
| 556.5 | 3 | 0.00 | N | Y |
| 580.4 | 1 | 0.00 | N | N |
| 581.4 | 3 | 0.02 | Y | N |
| 593.1 | 3 | 0.00 | N | Y |
| 624.0 | 3 | 18.35 | Y | N |
| 633.5 | 1 | 0.00 | N | Y |
| 635.5 | 2 | 0.00 | N | Y |
| 653.4 | 2 | 0.00 | N | N |
| 663.8 | 3 | 0.00 | N | Y |

[a]degeneracy; [b]infrared intensity (debye$^2$/Å$^2$/atomic mass unit),[c]infrared active,[d]raman active

Table 15. Computed vibrational modes at q=(0,0,0) for $YPO_4$

| $\nu(cm^{-1})$ | g[a] | IR Int[b] | IR[c] | RM[b] |
|---|---|---|---|---|
| 154.2 | 1 | 0.0 | N | N |
| 166.4 | 2 | 0.0 | N | Y |
| 190.3 | 1 | 0.0 | N | Y |
| 236.6 | 2 | 0.0 | N | Y |
| 237.7 | 2 | 10.1 | Y | N |
| 244.0 | 1 | 0.0 | N | N |
| 324.6 | 2 | 0.0 | N | Y |
| 339.2 | 1 | 0.0 | N | Y |
| 351.9 | 1 | 29.9 | Y | N |
| 352.9 | 1 | 0.0 | N | Y |
| 381.8 | 2 | 2.3 | Y | N |
| 418.0 | 1 | 0.0 | N | N |
| 518.9 | 1 | 0.0 | N | Y |
| 526.3 | 2 | 4.0 | Y | N |
| 577.5 | 1 | 0.0 | N | N |
| 584.8 | 2 | 0.0 | N | Y |
| 647.8 | 1 | 11.4 | Y | N |
| 669.2 | 1 | 0.0 | N | Y |
| 954.6 | 2 | 48.2 | Y | N |
| 957.6 | 1 | 0.0 | N | N |
| 971.7 | 1 | 0.0 | N | Y |
| 995.8 | 2 | 0.0 | N | Y |
| 1023.3 | 1 | 0.0 | N | Y |
| 1024.8 | 1 | 53.4 | Y | N |

[a]degeneracy; [b]infrared intensity (debye$^2$/Å$^2$/atomic mass unit),[c]infrared active,[d]raman active

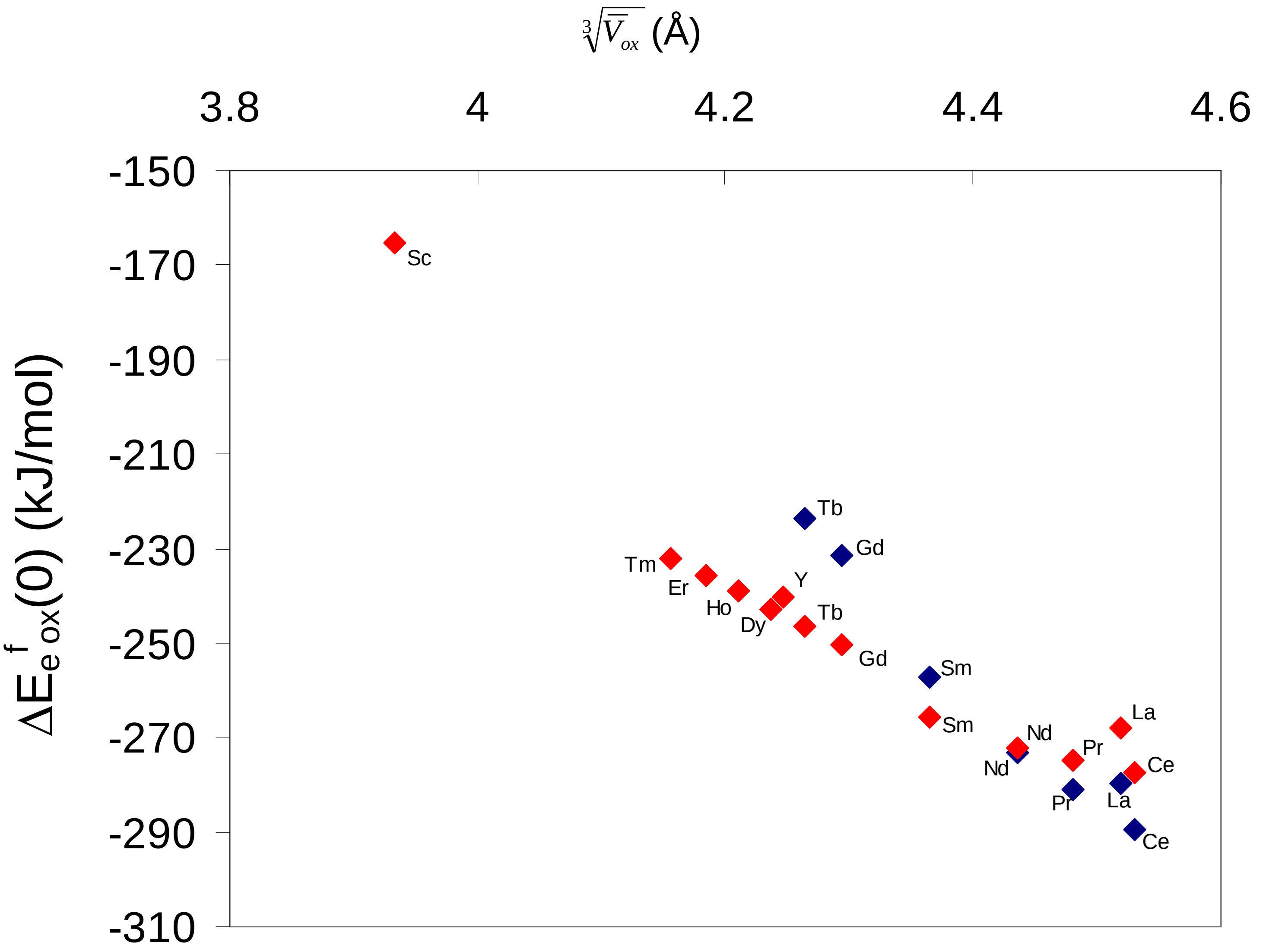

Figure 1. Correlation between $\Delta E_e{}^f{}_{ox}(0)$ and the cube root of the computed volume of the $RE_2O_3$ phase. (m) monazite, (x) xenotime

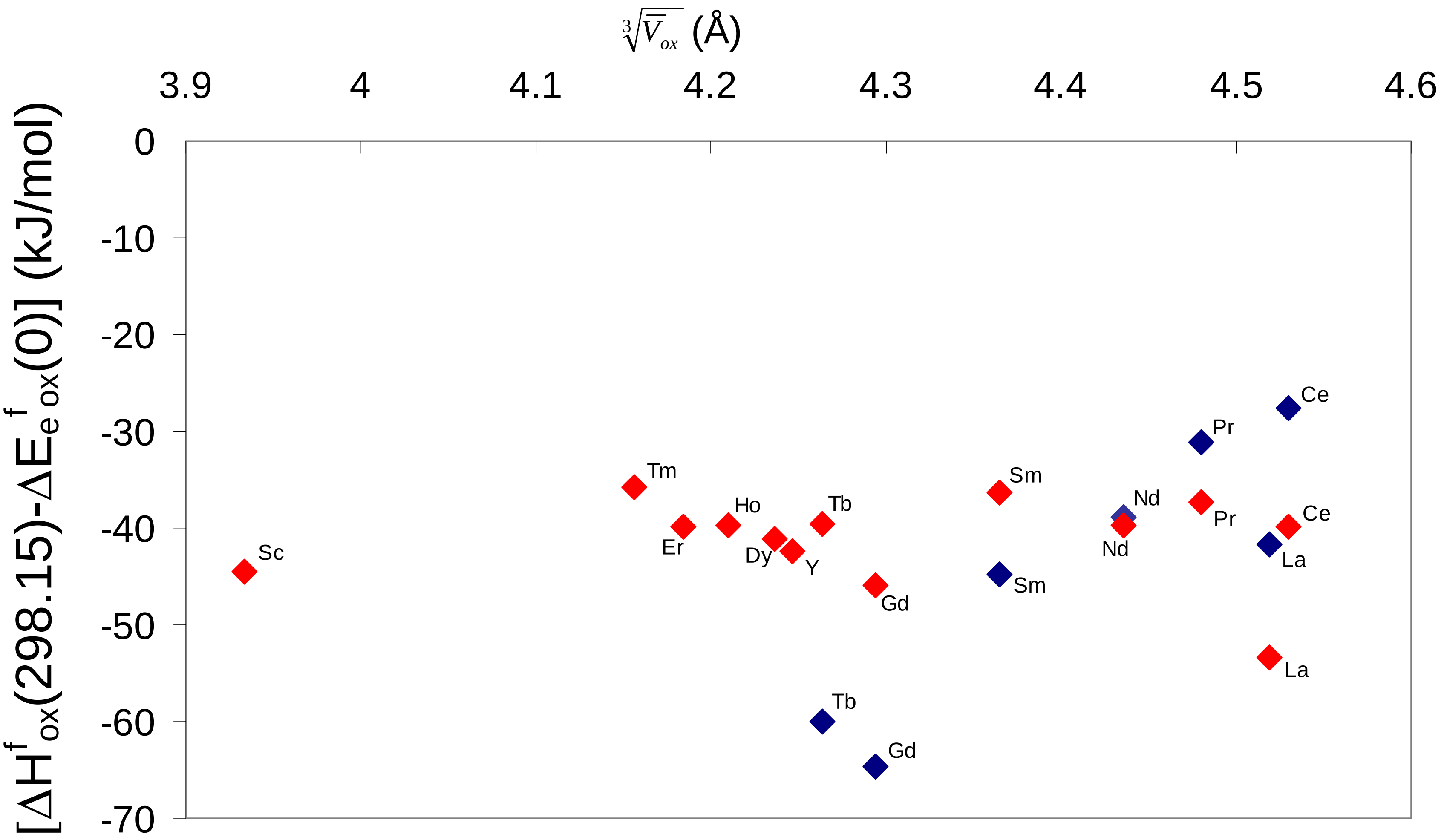

Figure 2. Correlation between the difference of the measured heat of formation and the computed electronic energy $[\Delta H^f_{ox}(298.15)-\Delta E_e{}^f_{ox}(0)]$ and the cube root of the computed volume of the cubic $RE_2O_3$.

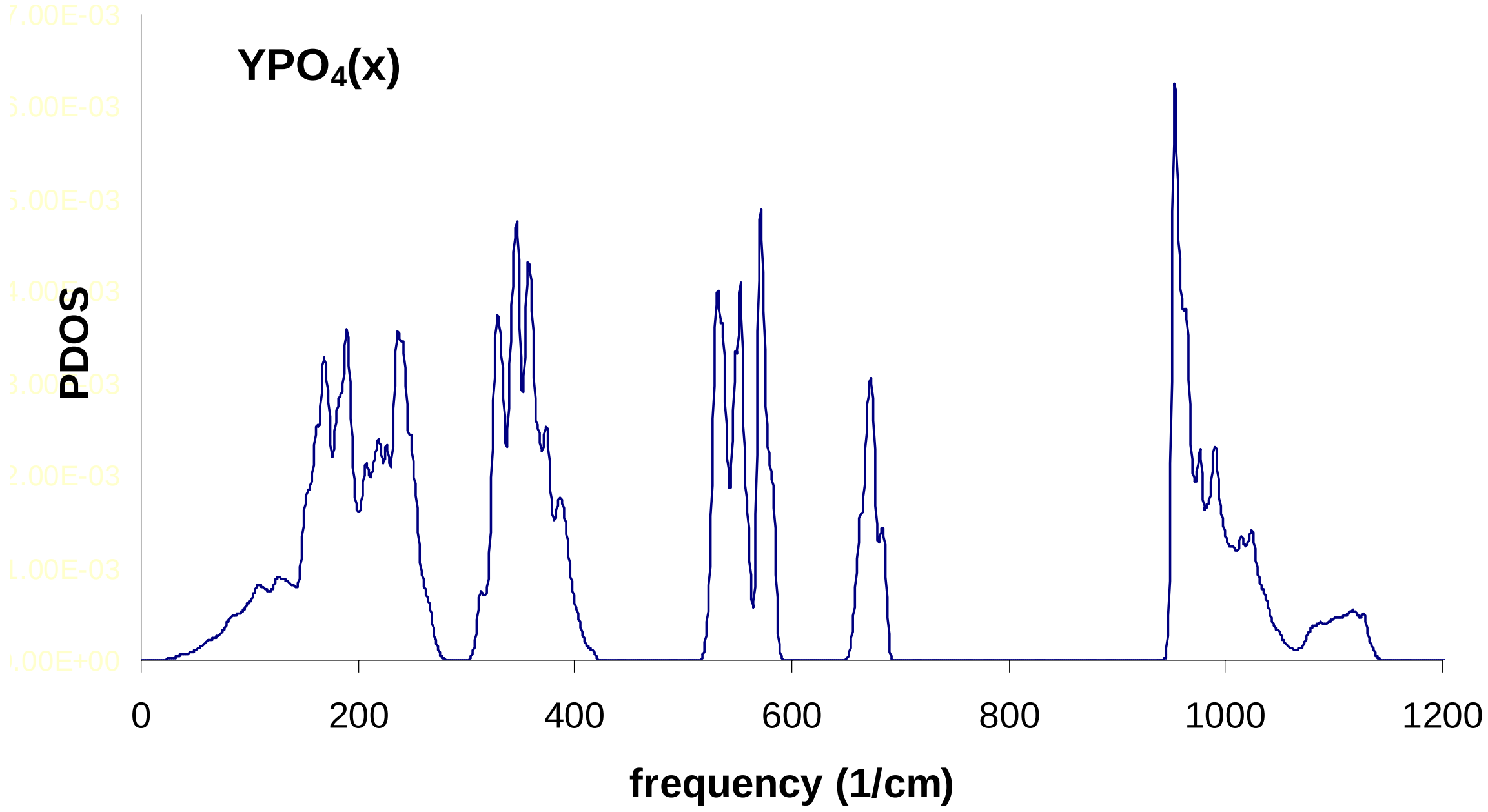

Figure 3. Phonon density of states computed for $YPO_4$-xenotime.

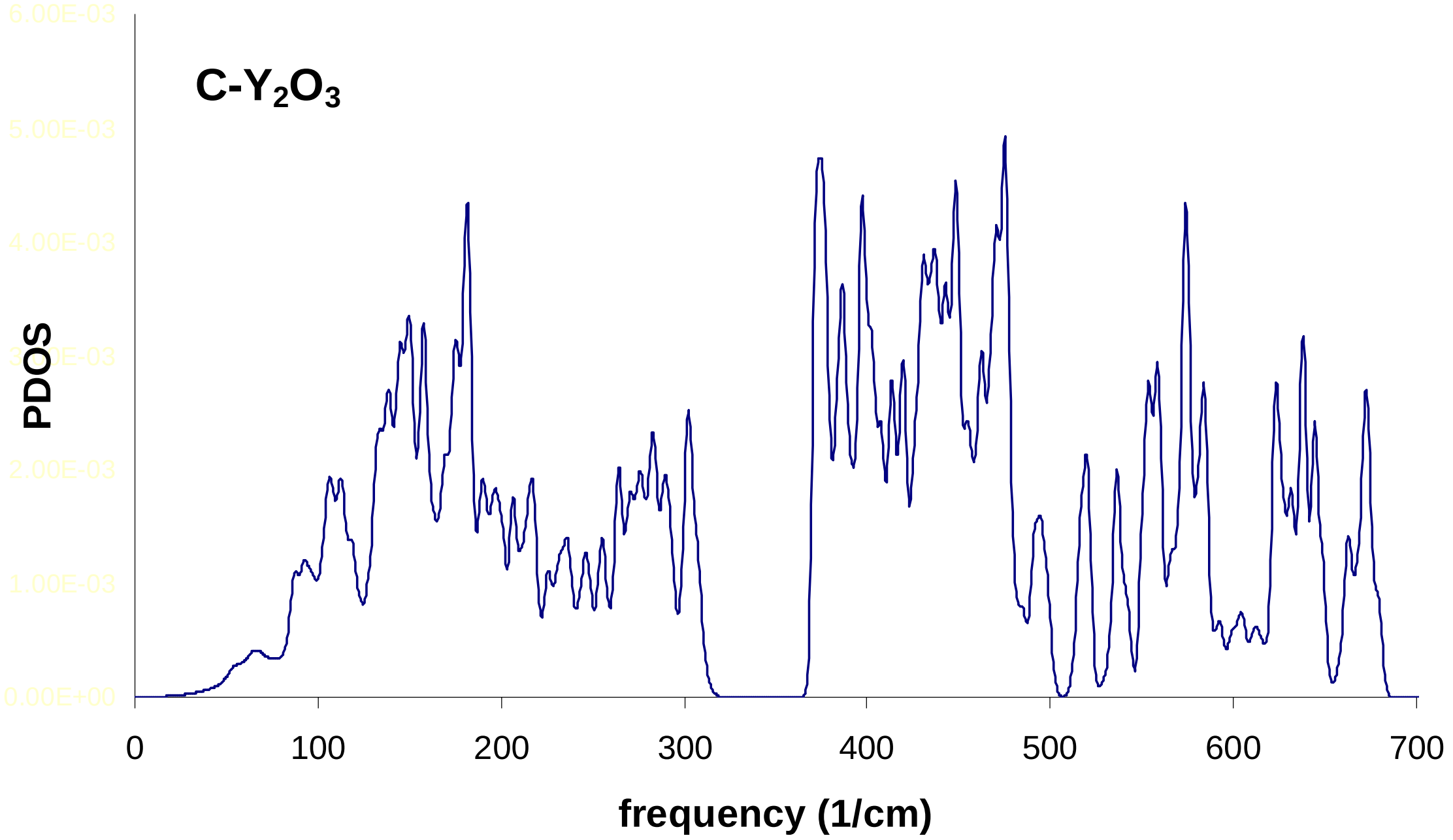

Figure 4. Phonon density of states computed for cubic $Y_2O_3$.

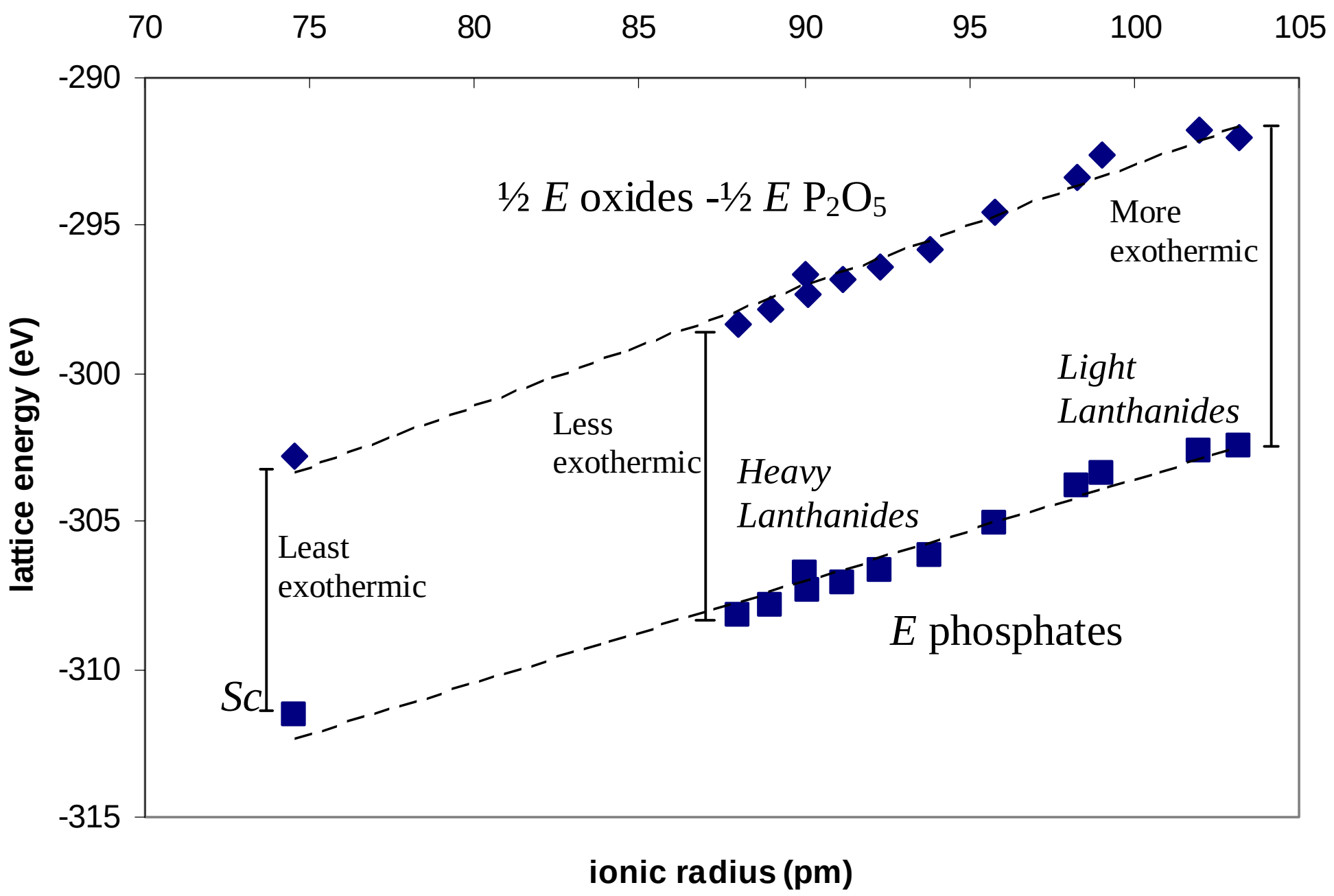

Figure 5. Electrostatic lattice energies E as a function of ionic radius for RE oxide and phosphate phases. The lattice energy of *h*-$P_2O_5$ has been added to the oxides and the sum multiplied by ½.

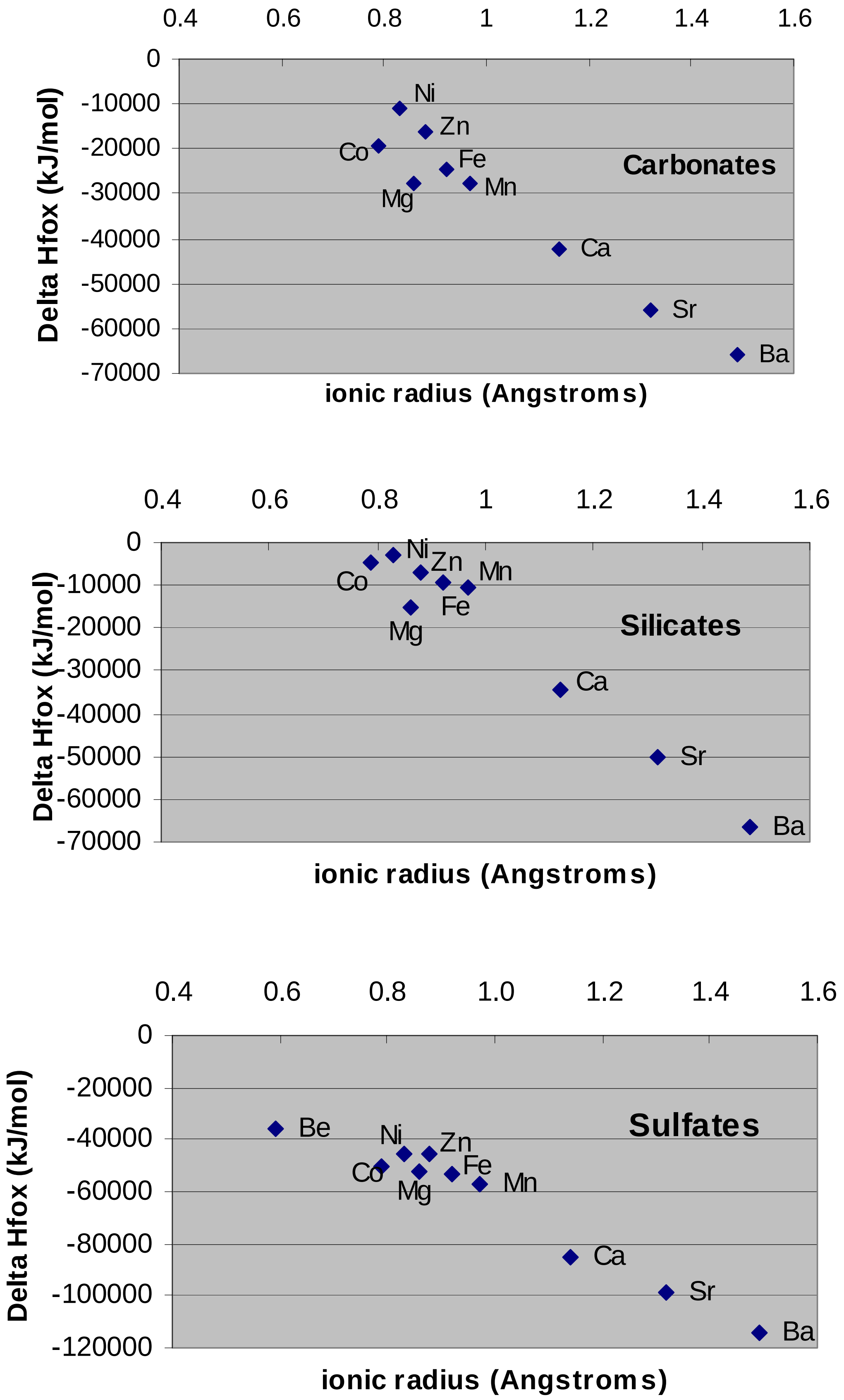

Figure 6. Correlation between heat of formation from the oxides and ionic radius for carbonate ($MCO_3$), orthosilicate ($M_2SiO_4$), and sulfate($MSO_4$) compounds with divalent anions. Data taken from FACTSAGE thermodynamic database.